%% file: main.tex
\documentclass[sigconf,screen]{acmart}

\usepackage{booktabs}
\usepackage{algorithm}
\usepackage{algorithmic}
\usepackage{amsmath}
\usepackage{siunitx}
\usepackage{graphicx}
\usepackage{array}
\usepackage{tabularx}
\usepackage{graphicx}
\usepackage{subcaption}
\usepackage[table]{xcolor}
\usepackage{multirow} 
\usepackage{booktabs}
\usepackage{enumitem}

\acmYear{2026}\copyrightyear{2026}
\setcopyright{cc}
\setcctype[4.0]{by}
\acmConference[FSE Companion '26]{34th ACM Joint European Software Engineering Conference and Symposium on the Foundations of Software Engineering}{July 5--9, 2026}{Montreal, QC, Canada}
\acmBooktitle{34th ACM Joint European Software Engineering Conference and Symposium on the Foundations of Software Engineering (FSE Companion '26), July 5--9, 2026, Montreal, QC, Canada}
\acmDOI{10.1145/3803437.3805216}
\begin{document}

\title{From Task to Tutorial: An Automated GUI Framework for Excel Tutorial Document and Video Creation}

\author{Yuhang Xie}
\affiliation{%
  \institution{Peking University}
  \city{Beijing}
  \country{China}}
\email{yuhangxie@stu.pku.edu.cn}

\author{Jian Mu}
\affiliation{%
  \institution{Nanjing University}
  \city{Nanjing}
  \country{China}}
\email{mujian@smail.nju.edu.cn}

\author{Xiaojun Ma}
\affiliation{%
  \institution{Microsoft}
  \city{Beijing}
  \country{China}}
\email{xiaojunma@microsoft.com}

\author{Chaoyun Zhang}
\affiliation{%
  \institution{Microsoft}
  \city{Beijing}
  \country{China}}
\email{chaoyun.zhang@microsoft.com}

\author{Lu Wang}
\affiliation{%
  \institution{Microsoft}
  \city{Beijing}
  \country{China}}
\email{wlu@microsoft.com}

\author{Mengyu Zhou}
\affiliation{%
  \institution{Microsoft}
  \city{Beijing}
  \country{China}}
\email{mengyu.chou@gmail.com}

\author{Mugeng Liu}
\affiliation{%
  \institution{Peking University}
  \city{Beijing}
  \country{China}}
\email{lmg@pku.edu.cn}

\author{Si Qin}
\affiliation{%
  \institution{Microsoft}
  \city{Beijing}
  \country{China}}
\email{Si.Qin@microsoft.com}

\author{Qingwei Lin}
\affiliation{%
  \institution{Microsoft}
  \city{Beijing}
  \country{China}}
\email{qlin@microsoft.com}

\author{Saravan Rajmohan}
\affiliation{%
  \institution{Microsoft}
  \city{Redmond}
  \state{Washington}
  \country{USA}}
\email{saravan.rajmohan@microsoft.com}

\author{Shi Han}
\affiliation{%
  \institution{Microsoft}
  \city{Beijing}
  \country{China}}
\email{shihan@microsoft.com}

\author{Dongmei Zhang}
\affiliation{%
  \institution{Microsoft}
  \city{Beijing}
  \country{China}}
\email{dongmeiz@microsoft.com}

\renewcommand{\shortauthors}{Yuhang Xie et al.}

\input{sections/00_abstract}

\keywords{Excel Tutorials, Automatic Generation, MLLM Agent, Task-to-Tutorial Framework}


% \keywords{Excel Tutorials, Automatic Generation, MLLM Agent, Task-to-Tutorial Framework}


\maketitle

\input{sections/01_introduction}

\input{sections/02_related_work}
\input{sections/03_approach}
\input{sections/04_experiments}

\input{sections/06_discussion}
\input{sections/07_conclusion}

\bibliographystyle{ACM-Reference-Format}
\bibliography{main}

\end{document}

%% file: sections/00_abstract.tex
\begin{abstract}

% 1. Excel complex
% 2. rely on human
% 3. Save how much effort

% v2

Excel is one of the most widely used productivity tools across domains, offering rich functionality but also overwhelming users with its complexity. This creates a persistent demand for tutorials to support effective usage. However, while building and maintaining the Microsoft tutorial corpus, we observed that existing tutorials are manually created by experts, need frequent updates with each software release, and involve substantial human labor. Moreover, prior work has not achieved fully automated tutorial generation. In this paper, we present the first framework for automatically generating Excel tutorials directly from natural language task descriptions. Our framework first instantiates the task. Then a central component of this framework, Execution Agent, plans and executes the solution in Excel, and collects the intermediate artifacts required for tutorial construction. These artifacts are then transformed into both structured Excel documents and video demonstrations. To build a comprehensive tutorial corpus, we collected 1,559 task descriptions from real-world scenarios. In addition, we designed a systematic evaluation framework that integrates assessments from both large language models (LLMs) and human reviewers. Experimental results show that our framework improves task execution success rates by 8.5\% over state-of-the-art baselines. Moreover, the generated tutorials demonstrate superior readability and instructional effectiveness, often approaching or surpassing expert-authored materials. Importantly, the automated pipeline eliminates manual labor and reduces time costs to 1/20 of expert authoring, making scalable and high-quality tutorial generation practical for the first time. 
% Furthermore, our framework generalizes well and can be easily extended to other software applications.

\end{abstract}

\begin{CCSXML}
<ccs2012>
   <concept>
       <concept_id>10011007.10011074.10011111.10010913</concept_id>
       <concept_desc>Software and its engineering~Documentation</concept_desc>
       <concept_significance>500</concept_significance>
       </concept>
   <concept>
       <concept_id>10003120.10003121.10003124.10010865</concept_id>
       <concept_desc>Human-centered computing~Graphical user interfaces</concept_desc>
       <concept_significance>300</concept_significance>
       </concept>
   <concept>
       <concept_id>10010405.10010497.10010510.10010515</concept_id>
       <concept_desc>Applied computing~Multi / mixed media creation</concept_desc>
       <concept_significance>500</concept_significance>
       </concept>
 </ccs2012>
\end{CCSXML}

\ccsdesc[500]{Applied computing~Multi / mixed media creation}
\ccsdesc[500]{Software and its engineering~Documentation}
\ccsdesc[300]{Human-centered computing~Graphical user interfaces}

%% file: sections/01_introduction.tex
\section{Introduction}

Excel provides a comprehensive set of functionalities and is extensively adopted across domains such as finance, analytics, and scientific research~\cite{chan1996use,hacker2017financial,powell2019business}. 
Its versatility makes it a critical tool for data management, statistical analysis, and visualization. 
At the same time, this richness of features creates significant learning barriers, as many users struggle to identify the appropriate operations or to integrate multiple functions effectively~\cite{carroll1990nurnberg,ko2007information}. 
To mitigate these challenges, tutorials have long served as an essential support mechanism, guiding users through common tasks and enabling more effective utilization of Excel’s capabilities. However, through Microsoft’s operational experience and long-term product maintenance, we observe that the prevailing approach of manually authoring tutorials by domain experts incurs substantial labor costs and limits coverage across the diverse range of tasks that users encounter~\cite{grossman2010toolclips}. Moreover, as software evolves, these tutorials must be continuously updated: outdated screenshots, obsolete descriptions, and inconsistent workflows reduce their instructional value~\cite{liu2024having}. 
This dependence on manual authoring highlights the need for a scalable, automated approach to tutorial generation.

Prior research has explored semi-automated methods for tutorial creation. 
Some approaches generate tutorials based on expert demonstrations~\cite{chi2012mixt,grabler2009generating}, while others transform existing manuals into instructional videos or annotated guides~\cite{liu2024having,zhong2021helpviz}. 
Although these methods improve efficiency, they fundamentally rely on curated expert solutions, still requiring substantial human effort and preventing a fully end-to-end pipeline.
Recent advances in LLMs and agent systems make automated tutorial generation increasingly feasible~\cite{zhang2024ufo,liu2025infigui}. 
Spreadsheet agents can already automate a variety of operations~\cite{li2023sheetcopilot}, but their code-driven execution remains opaque to end users, offering limited pedagogical benefit. 
Similarly, Computer-Using Agents (CUAs) ~\cite{openai2025computer,zhang2025ufo2} demonstrate the ability to plan application-level tasks through simulated human interactions. 
However, applying such systems to Excel is uniquely challenging due to its densely structured interface and fine-grained operations (e.g., cell and border editing). 
Generating tutorials from task descriptions further complicates the task: the system must not only plan and execute tasks but also collect execution traces, screenshots, and contextual information suitable for tutorial construction. 
To date, no existing work addresses this end-to-end requirement.

In this paper, we present the first framework for automatically generating Excel tutorials directly from natural language task descriptions. Our framework first instantiates the task. Then the Execution Agent plans and executes the solution in Excel, and collects the intermediate artifacts required for tutorial construction. These artifacts are subsequently transformed into step-by-step instructional documents and narrated video demonstrations. 
Different tutorial formats offer complementary learning benefits: documents allow for efficient skimming of procedures, while videos provide clear demonstrations of fine-grained execution steps and are particularly useful for users who prefer visual or auditory guidance~\cite{torrey2007pages}. 

To support development and evaluation, we curate a dataset of 1,559 real-world Excel tasks spanning 28 operations and 17 target object categories. 
This dataset covers Excel-level tasks that are absent from prior work and enables the generation of comprehensive tutorials at scale. 
We further design a systematic evaluation protocol that combines human expert judgments with LLM-based assessments. 
Experimental results show that our framework improves execution success rates by 8.5\% over state-of-the-art baselines. 
Moreover, the generated tutorials demonstrate superior readability and instructional effectiveness, often approaching or surpassing expert-authored materials, while reducing authoring costs to one-twentieth of manual effort. Furthermore, experiments on Word and PowerPoint datasets indicate that our framework generalizes well and can be easily extended to other applications.
Our main contributions are as follows:
\begin{itemize}[leftmargin=*]
    \item We propose the first end-to-end framework for generating Excel tutorials directly from natural language task descriptions, producing both instructional documents and video demonstrations. 
    \item We introduce an Execution Agent that plans and executes Excel tasks while collecting intermediate artifacts for tutorial generation, achieving a task execution success rate of 39.58\% and surpassing state-of-the-art baselines. 
    \item A dataset of 1,559 real-world Excel tasks is curated, covering a wide range of operations and target object categories to enable scalable and representative tutorial generation. 
    % demonstrating
    \item A systematic evaluation protocol is designed with both human and LLM-based judges, showing that the generated tutorials are effective, readable, and substantially reduce manual labor costs.
    \item We deploy our framework as a production-grade tutorial generation framework in Microsoft M365 Excel. In a real production setting, it generated 1,400 Excel tutorials, reducing content authoring time by over 99\% compared to manual writing—while achieving a 9\% improvement in user engagement after being released to real users.
\end{itemize}

%% file: sections/02_related_work.tex
\section{Related Work}

\subsection{Automated Tutorial Generation}
Prior work has explored semi-automated tutorial generation using pre-existing materials such as user demonstrations
and instructional videos. MixT \cite{chi2012mixt} produces mixed-media tutorials by combining static instructions with video segments derived from user demonstrations. Similarly, an approach~\cite{truong2021automatic} has been proposed to automatically construct hierarchical tutorials from makeup instructional videos.
Other research \cite{grabler2009generating} captures software operations along with screencast recordings and converts them into document-style tutorials enriched with text descriptions and annotated step images. In the domain of physical tasks, some approaches \cite{denning2011meshflow,grossman2010chronicle} produce instructional videos through semi-automatic editing of creator-annotated demonstrations. Additionally, several studies have explored transforming static textual content (e.g., user manuals) into richer, multimedia-based tutorials.
HelpViz \cite{zhong2021helpviz} transforms text-based instructions into contextual visual guides via action parsing and interaction simulation, and systems such as M2V \cite{liu2024having} and HowToCut \cite{chi2021automatic} generate instructional videos from manuals or Markdown tutorials. While these methods produce guidance, they all rely on pre-existing instructional materials and do not achieve fully automated tutorial generation directly from task descriptions.

\subsection{Spreadsheet Agents}

\begin{figure*}[htbp]   
\vspace{-10pt}
  \centering              
  \includegraphics[width=0.75\textwidth]{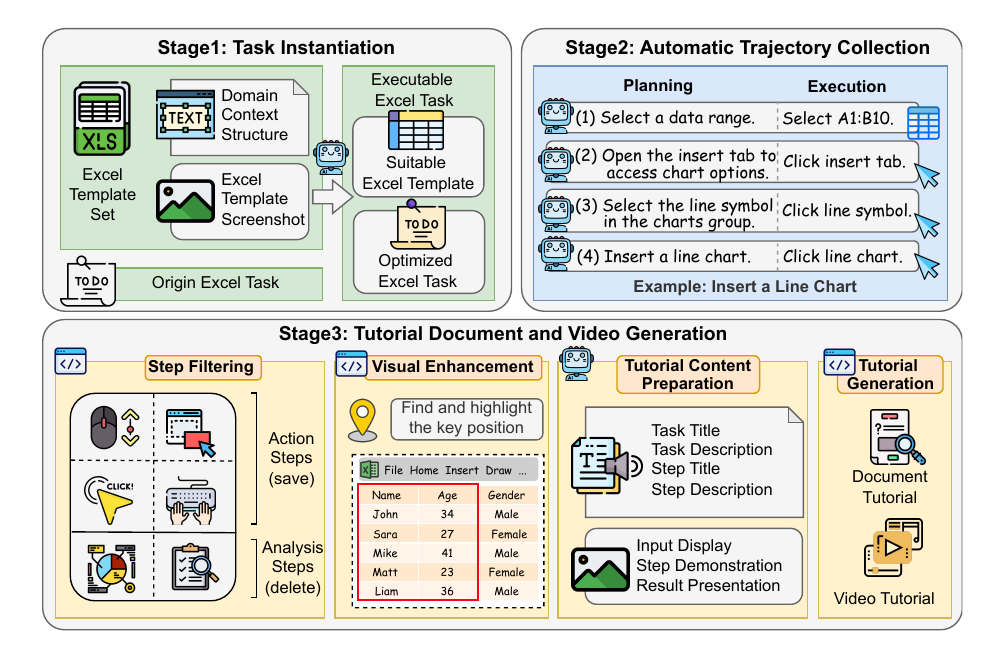} 
  \vspace{-18pt}
  \caption{Automated Workflow for Excel Task Execution and Tutorial Generation.}  
  \label{fig:overview}   
  \vspace{-8pt}
\end{figure*}

With rapid LLM advances, recent spreadsheet agents increasingly leverage LLM capabilities to tackle spreadsheet tasks. SheetCopilot \cite{li2023sheetcopilot} enables stable interaction between LLMs and spreadsheets by defining atomic actions within a state machine-based task planning framework. SheetAgent \cite{chen2025sheetagent} introduces a modular Planner-Informer-Retriever architecture that effectively addresses long-horizon spreadsheet tasks through iterative reasoning. SheetMind \cite{zhu2025sheetmind} builds a multi-agent system with defined roles, enhancing execution reliability. Despite progress in automating table-level tasks, existing agents still struggle with complex scenarios \cite{li2023sheetcopilot,zhu2025sheetmind,wang2025odysseybench}. Moreover, their code–based operations are opaque to end users and fail to provide interface-based, interpretable learning paths.

\vspace{-3pt}
\subsection{Computer-Using Agents (CUAs)}
Recent advances in desktop automation have led to various LLM-powered agent systems. UFO \cite{zhang2024ufo} represents one of the multi-agent automation frameworks for Windows, emphasizing GUI interaction through the integration of UI Automation and visual perception. NAVI \cite{bonatti2024windows}, a single-agent baseline from WAA, leverages both screenshots and accessibility metadata to facilitate GUI understanding. OmniAgent \cite{lu2024omniparser} combines OmniParser for visual grounding with GPT-based action planning, enabling robust multimodal reasoning. Agent S \cite{agashe2024agent} employs a multi-agent architecture with experience-driven hierarchical planning, optimized for executing complex, multi-step tasks. Operator \cite{openai2025computer}, a recent high-performance CUA from OpenAI, simulates human-like mouse and keyboard operations based on screenshot inputs. Despite their promise, these systems often exhibit shallow OS integration and heavy reliance on brittle visual inputs, resulting in suboptimal accuracy on fine-grained tasks such as spreadsheet operations \cite{zhang2024ufo,chen2025map}.

\vspace{-4pt}
\section{Task Definition}

In this work, we focus on the automatic generation of tutorials for Excel tasks, producing outputs in two complementary formats: documents and videos. We first introduce the characteristics of these tutorial formats, then formally define the problem setting.

\vspace{-4pt}
\subsection{Tutorial Formats} 
Textual tutorials provide structured step sequences with natural language instructions and optional illustrative images~\cite{zhong2021helpviz}, enabling efficient scanning and reference. They can be delivered in formats like HTML or Markdown, supporting diverse instructional contexts. Demonstration videos combine screen recordings with audio narration~\cite{torrey2007pages,tuncer2020pause}, presenting operations dynamically while conveying the overall task workflow~\cite{chi2012mixt,truong2021automatic,zhu2021gif}. Voice narration aids learners preferring auditory guidance, making videos an effective complement to textual tutorials.

\vspace{-4pt}
\subsection{Problem Definition} 
The \emph{task-to-tutorial problem} transforms a natural language task description (e.g., ``create a bar chart'') into a tutorial package containing both text and video. Each tutorial consists of an ordered sequence of steps, each with: step index, step title, step description, and supporting materials (screenshots or audio). Unlike prior methods, our system does not rely on pre-existing demonstrations. It autonomously plans and executes tasks in Excel, captures execution traces, and generates instructional materials, integrating task completion and tutorial construction. The goal is to produce tutorials that are accurate, pedagogically effective, scalable, and minimize manual authoring effort.

%% file: sections/03_approach.tex
\vspace{-4pt}
\section{Approach}
% done: add datasets here.
% \fix{Dataset collection is described at the beginning of the paragraph, as it is not part of the tutorial generation process. In the task definition, the task itself serves as the input; therefore, task collection should not be considered part of the generation workflow?}\cz{Why dataset is not in the stage?}

% In this section, we introduce data collection process and the overall pipeline for Excel tutorial generation as \autoref{fig:overview}.

We propose \emph{an automated GUI framework} for Excel tutorial generation that addresses the \emph{task-to-tutorial} problem, as illustrated in \autoref{fig:overview}. 
This framework enables the creation of both instructional documents and video tutorials directly from natural language task descriptions. 
It is supported by a curated dataset of 1,559 real-world Excel tasks, which provides the foundation for training and evaluation. 
The framework itself consists of three stages. 
\emph{Stage 1: Task Instantiation} transforms a raw task description into an executable Excel task by aligning it with appropriate templates and contextual structures. 
\emph{Stage 2: Automatic Trajectory Collection} employs the Execution Agent to plan and execute the task while recording a fine-grained sequence of user-like actions (e.g., selecting ranges, navigating menus, applying formulas). 
\emph{Stage 3: Tutorial Document and Video Generation} converts the collected trajectory into user-facing materials by filtering essential steps, enhancing visual clarity, and preparing step-level descriptions. 
The final outputs are both structured documents and narrated video tutorials, offering complementary modalities for effective learning.
In the following subsections, we first introduce the dataset construction process and then describe each of the three stages in detail.
 
% (1) \textbf{Task Instantiation} (\autoref{sec:task_instantiation}), aiming at generating executable task instances through template matching and task refinement;
% (2) \textbf{Solution Planning and Execution} (\autoref{sec:solution-planning-execution}), focused on planning and executing the operation sequence for the task via an Excel task executor; 
% (3) \textbf{Tutorial Document and Video Generation} (\autoref{sec:tutorial-generation}), which generates structured tutorial components based on operation step sequences and synthesizes them into video and document tutorials. 
% Below, we provide detailed information for each stage of the EGA.

\vspace{-3pt}
\subsection{Dataset Construction} 
\label{subsec:dataset contruction}
%%chaoyun:我们需要什么样的query，是true query，不是生成的。还有高频但是目前的dataset不是很满足。所以这motivate我们collect新的q

Existing datasets for Excel are derived from  either manually crafted examples ~\cite{payan2023instructexcel} or Excel exam datasets~\cite{chen2025sheetagent}. These tasks are artificially generated and intended for examination testing. As such, they do not correspond to authentic user behavior patterns.
Other datasets collect queries from Excel forums where users seek help online ~\cite{li2023sheetcopilot, ma2024spreadsheetbench}. These queries are formulated under constrained and highly specific conditions, thereby limiting their generalization to broader user scenarios.
Moreover, these existing datasets ~\cite{payan2023instructexcel, chen2025sheetagent, li2023sheetcopilot, ma2024spreadsheetbench} focus exclusively on spreadsheet-level operations and omit those at the Excel application level. We define Excel-level tasks as operations that target the Excel application itself (e.g., adding the Camera tool, opening the navigation pane), rather than the specific spreadsheet content. In contrast, spreadsheet-level tasks directly affect the content, formatting, structure, or other elements within a particular spreadsheet (e.g., inserting a new row, merging cells).
% \fix{finish}\cz{Like what? These dataset are also not realistic. We need to first highlight that need to collect real-world and representative data for tutorial making but current dataset may not satisfied. Highlight they are real-world}

To obtain real-world Excel tasks that are pedagogically meaningful, broadly transferable, and diverse in scope, we collect task descriptions from three primary sources:
\begin{itemize}[leftmargin=*]
    \item \textbf{Search Queries (Search)}: We extract task descriptions from Bing Search queries related to Excel operations, which reflect real user needs and common practical challenges.
    \item \textbf{Community Websites (Websites)}: We crawl community-driven websites, forums, and technical Q\&A platforms (e.g., WikiHow) to obtain task descriptions. Such sources provide a wide range of authentic and diverse real-world tasks.
    \item \textbf{In-App Help Content (In-App)}: We mine task descriptions from official Office documentation and tutorials, which cover a broad set of standardized operations.
\end{itemize}

The diverse composition of sources ensures comprehensive coverage of practical user needs, authentic interaction scenarios, and standardized documentation practices. After filtering the collected data for semantic ambiguity using GPT-5 and removing duplicates based on embedding similarity, we obtained the original task set containing 1,559 entries.

\begin{table}[t]
% \vspace{-8pt}
% \vspace{-2pt}
\centering
\caption{Dataset Statistic.}
\vspace{-10pt}
\label{tab:dataset}
{
\renewcommand{\arraystretch}{0.96} 
\begin{tabular}{lccc}
\toprule
\textbf{Dataset} & \textbf{Count} & \textbf{Spreadsheet-Level (\%)} & \textbf{Excel-Level (\%)} \\
\midrule
 Search & 700 & 87.71 & 12.29 \\
Websites & 721 & 75.59 & 24.41 \\
In-App & 138 & 82.61 & 17.39 \\
\midrule
All & 1559 & 81.65 & 18.35 \\
\bottomrule
\end{tabular}
}
\vspace{-11pt}
\end{table}

% mugeng : 容易被argue
% \cz{Interpret the results}
To further investigate the coverage and representativeness of real-world conditions, we analyze the involved objects (e.g., chart, table) and operations (e.g., add, delete) in our dataset. Specifically, we employ GPT-5 to extract the objects and operations in tasks, and further identify whether the task is Spreadsheet-level or Excel-level (see ~\autoref{tab:dataset}). The classification results
% (see ~\autoref{fig:type_object_distribution}) 
show that the dataset comprises 28 operation categories and 17 target object categories. Our dataset exhibits substantial diversity, with many distinct combinations of operation categories and target object categories, which is attributable to its broad and authentic data sources.

\vspace{-6pt}
\subsection{Stage 1: Task Instantiation}
\label{sec:task_instantiation}

% ambiguous，加一个图或者加一个例子。like 加一个symbol，需要做一个实例化。
% Template match 加一个图
% \cz{This section needs a figure, like Figure 4 in \url{https://arxiv.org/pdf/2412.10047}}

To produce Excel tutorials, we need real Excel environments to perform the tasks.
In \autoref{subsec:dataset contruction}, real queries are collected without the matching Excel files. Furthermore, raw real-world queries are sometimes ambiguous. In this stage, we ensure that the task is clear and feasible. For example, the original collected query ``add a title'' is ambiguous, because it is unclear where the title should be added. In contrast, the instruction ``add the title `AI Agent' to the table in range A1:G3'' constitutes an actionable and clear task.
Therefore, we synthesize instantiated tasks from the original tasks through the following three-stage process.
% : create template set, select suitable template, and optimize task description.

% \fix{finish}\cz{Need to mention that these queries are prototyping (give an example) and not grounded in an environment.}

\textbf{Template Set Creation.} 
To provide an execution environment for each task, we constructed a template set consisting of sufficient Excel files with diverse domains and compositions. Each template includes an Excel file, screenshots, and a description of the file content and data structure.

\textbf{Template Matching.}  
The execution of Excel tasks requires a compatible Excel environment. For example, task ``add a title for the chart'' should be executed on an Excel file that contains charts. In detail, we leverage LLMs to analyze the essential elements for each task and automatically match the task to a suitable Excel file. 

\textbf{ Executable Task Construction.}  
In this step, we rewrite the ambiguous queries into clear, executable instructions aligned with the matched Excel template.
For example, a vague request like ``sort the data'' is rewritten as ``sort the data by age in ascending order''. We use LLMs to identify the specific objects in the template to which the operation applies and rewrite the instructions to be concrete.

\vspace{-3pt}
\subsection{Stage 2: Automatic Trajectory Collection}
\label{sec:solution-planning-execution}

\begin{figure*}[t]  
  \vspace{-11pt}
  \centering              
  \includegraphics[width=0.92\textwidth]{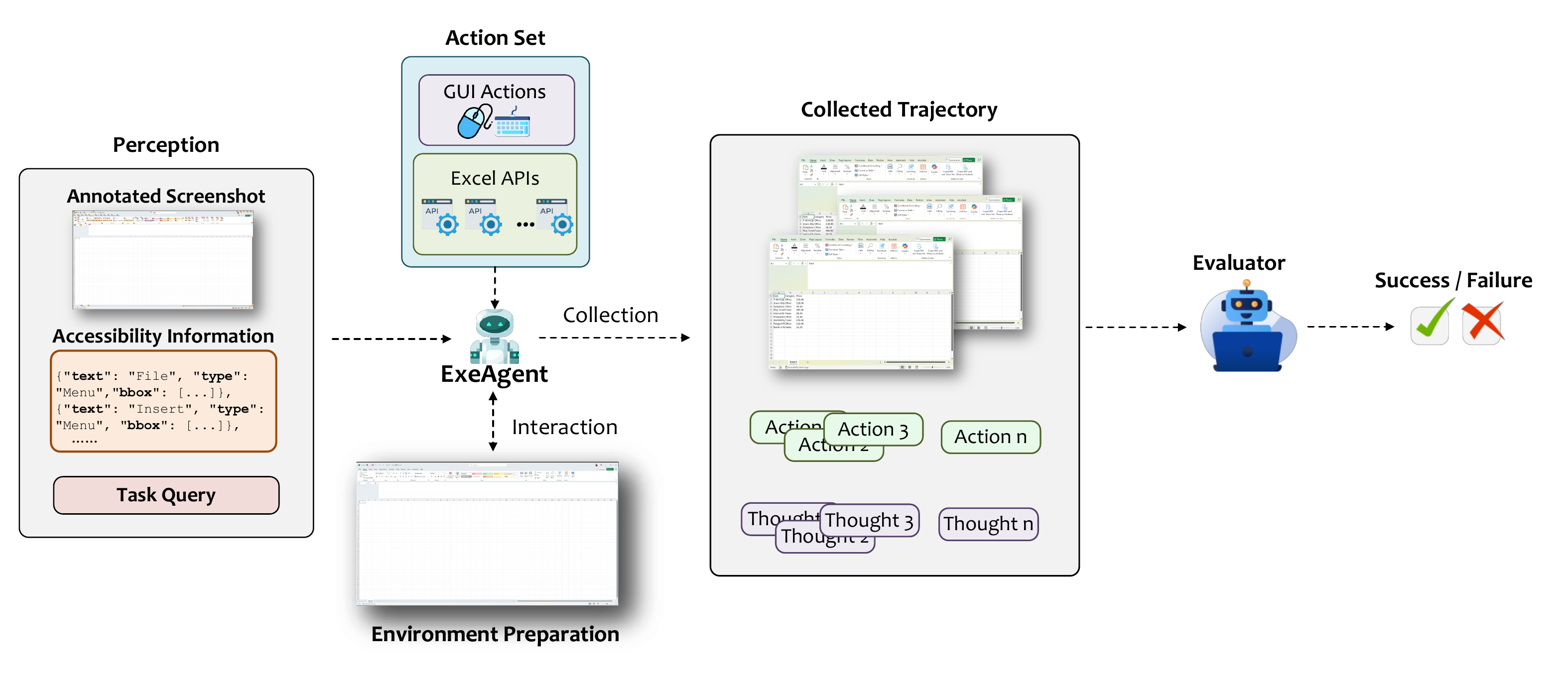} 
  \vspace{-15pt}
  \caption{An overview of the trajectories collection workflow with ExeAgent.}  
  \label{fig:data_collection}   
  \vspace{-10pt}
\end{figure*}

% \begin{table*}[t]
% \vspace{-8pt}
% \centering
% \caption{Examples of Excel-related APIs incorporated in the ExeAgent.}
% \vspace{-8pt}
% \begin{tabularx}{\textwidth}{l|X}
% \hline
% \textbf{API Name} & \textbf{Description} \\
% \hline
% table2markdown & Converts a table from a given Excel sheet into markdown format. \\
% \hline
% insert\_excel\_table & Inserts a set of data into a specified location in an Excel sheet as a table. \\
% \hline
% select\_table\_range & Select a specified range of cells in an Excel sheet. \\
% \hline
% % set\_cell\_value & Sets the value or formula of a given cell in an Excel sheet. \\
% % \hline
% auto\_fill & Automatically fills a range of cells based on recognized patterns. \\
% \hline
% reorder\_columns & Reorders the columns of an Excel sheet according to a specified order. \\
% \hline
% \end{tabularx}
% \label{tab:excel_functions}
% \vspace{-10pt}
% \end{table*}

Given the instantiated queries, the action plans, execution trajectories, and the execution results are produced in this stage. Traditionally, this process requires human effort for task execution and log collection, which is costly and challenging to scale up \cite{wang2024large}.

To address this challenge, we design a specialized execution agent, ExeAgent, to automatically complete Excel queries in batch (see \autoref{fig:data_collection}). Before execution, ExeAgent prepares the environment and performs  the task execution. The complete execution trajectory is recorded and subsequently evaluated by an LLM-based evaluator, which automatically determines task completion. Only successfully completed trajectories are retained for tutorial generation. After execution, ExeAgent closes the template files to reset the environment and prepare for the next task. Notably, the entire workflow is \emph{fully automated} and requires no human intervention.

\vspace{-3pt}
\subsubsection{Executor Design in ExeAgent}
The core component of ExeAgent is its \emph{executor}, as the executor’s success rate and the quality of its execution traces directly determine the amount and usefulness of the generated tutorials. The executor is designed as a GUI-based agent because such agents mimic natural human interactions with the interface through mouse clicks and keyboard inputs \cite{zhang2024large}. This design ensures that the collected trajectories are intuitive and closely aligned with real user behavior, which makes them particularly suitable for tutorial generation.

However, existing pure-visual GUI-based Computer-Using Agents (CUAs) (e.g., \cite{anthropic2024, cua2025, zheng2025vem, wu2025gui}) are inadequate for handling complex Excel tasks for two reasons. First, Excel interfaces are highly intricate, with hundreds of cells arranged in diverse layouts and embedded diagrams. Relying solely on screenshot-based visual grounding is error-prone and often results in failed executions. Second, CUAs usually represent actions as screen coordinates $(x, y)$, which lack semantic meaning, making the trajectory difficult to interpret and unsuitable for generating tutorials. These limitations motivate the design of a specialized executor tailored to Excel automation and tutorial generation.

\vspace{-3pt}
\subsubsection{Accessibility-Augmented Perception.}
To address the challenges of screenshot-only grounding, ExeAgent integrates accessibility information into its perception pipeline. Accessibility metadata provide a structured list of actionable GUI elements, each annotated with its name, type, and bounding box. For example: \{"name": "File Menu", "type": "MenuItem", "bbox": [10, 20, 80, 40]\}. Such information is overlaid on screenshots using the Set-of-Marks (SoM) mechanism \cite{yang2023set}, thereby enriching visual perception with semantic context. This augmentation improves execution in two ways: (1) it reduces reliance on brittle screenshot understanding, particularly in Excel’s cell-dense interface, and (2) it enables semantically meaningful action representation (e.g., click on \texttt{File Menu}) instead of opaque coordinate-based actions. As a result, the trajectories become both more robust and interpretable for tutorial generation.

\vspace{-3pt}
\subsubsection{Hybrid GUI–API Execution.}
Although GUI actions such as mouse clicks and keystrokes resemble human behavior, they remain error-prone and may reduce execution success rates \cite{zhang2025api}. To mitigate this, ExeAgent employs a hybrid execution model that combines GUI actions with dedicated Excel APIs (e.g., table2markdown, which converts tables from Excel sheets into Markdown format). API calls leverage well-defined programmatic endpoints to perform operations precisely and efficiently, substantially reducing execution errors. In this design, GUI actions handle generic interactions not covered by APIs, while complex or fragile operations are offloaded to the API layer. For API-completed steps, we additionally collect information such as the corresponding operation regions, enabling the system to reconstruct the equivalent GUI actions during tutorial generation to facilitate user understanding. This hybrid approach ensures both generality and robustness in task execution.

% (see~\autoref{tab:excel_functions})

\vspace{-3pt}
\subsubsection{Execution Workflow.}
With accessibility-enhanced perception and a hybrid action set, ExeAgent follows a reactive execution framework. Upon receiving a query, it processes the screenshots and accessibility metadata, reasons through the task using a Chain-of-Thought approach~\cite{wei2022chain, dingeverything}, and selects the most appropriate action iteratively until it determines the task is complete. During execution, ExeAgent systematically records multi-modal artifacts for tutorial generation, organizing them into a step-by-step trajectory. Each step in the trajectory includes: (1) a screenshot of the interface, (2) the intermediate reasoning trace, (3) the concrete action executed and its result, (4) the interaction location or targeted GUI element. These rich multi-modal artifacts provide sufficient detail to generate high-quality, human-friendly tutorials.

\vspace{-3pt}
\subsubsection{Automatic Evaluation}
Finally, before generating tutorials, it is necessary to filter out unsuccessful trajectories. Traditional approaches—such as using hand-crafted scripts to generate oracles or relying on human evaluators, are both costly and difficult to scale. To address this challenge, we adopt an \emph{LLM-as-a-judge} approach, leveraging GPT-4.1 as the evaluator. The evaluator consumes the multi-modal execution trajectory as input and automatically determines whether the task has been successfully completed. Only trajectories deemed successful are retained for tutorial generation. A preliminary experiment on 200 samples demonstrates that the LLM-based evaluator achieves an agreement rate of 90.5\% with human evaluation, indicating that it is both reliable and scalable.

\vspace{-3pt}
\subsection{Stage 3: Tutorial Document and Video Generation}
\label{sec:tutorial-generation}

\subsubsection{Step Filtering}
Execution trajectories in \autoref{sec:solution-planning-execution} could sometimes be redundant and too detailed even if the task is successfully executed. For example, "select range A1:B2" could sometimes be repeated twice. To make the tutorials concise and clear, we remove repeated steps and those that do not involve actual operations, as such steps primarily serve analytical purposes like "look at this chart". 
%Non-operational step examples include annotation (capturing the current window and marking control elements for subsequent analysis) and summary (providing a descriptive overview of the interface based on screenshots or control lists). 
By filtering the step sequences, we get a structured action flow that retains the effective interactions with the Excel interface.

\vspace{-3pt}
\subsubsection{Visual Enhancement}
% Highlighting key information and core actions has been shown to improve tutorial readability and reduce cognitive load~\cite{van2013eight,brame2017effective}. 
At this stage, we obtain clear interaction sequences, but it is difficult for users to follow the steps only by screenshots and textual instructions because some key actions could be hard to follow when no obvious visual cues are present. Therefore, for each screenshot, we draw red bounding boxes over key regions (operation locations) to direct attention and enable rapid localization. To provide a realistic demonstration and avoid occlusion, we overlay a cursor icon at the bottom-right corner of the boxed region, aiding learner comprehension.

\vspace{-3pt}
\subsubsection{Tutorial Content Preparation} 
% \fix{finish}\cz{Use different name,}
Based on the task description, the filtered step sequence log, and the enhanced visualization, we generate the following textual content of the tutorial:
\vspace{-3pt}
\begin{itemize}[leftmargin=*]
    \item Task title, provides an immediate and intuitive understanding of the tutorial’s purpose.
    %~\cite{farkas1999logical,van2004four}.
    \item Task description, establishes explicit learning objectives with environment settings. %~\cite{kay2014developing,van2013eight}.
    \item Step titles, concisely summarize the content of each operation. Decomposing core instructional content into a series of explicit and coherent steps facilitates comprehension. % ~\cite{brame2017effective,renkl2005worked,kirschner2006minimal}.
    \item Step descriptions provide guidance on execution and explain each step’s purpose, helping users link individual actions to the overall conceptual framework and avoid cognitive leaps. %~\cite{kay2014developing,renkl2005worked}.
\end{itemize}
\vspace{-3pt}
We generate textual content for document and video tutorials separately, as their linguistic and structural requirements differ. Specifically, step descriptions in document tutorials adopt a formal written style, whereas those in video tutorials are phrased as spoken narration, which is subsequently synthesized into audio. To ensure the generation of high-quality instructional content, we design a structured prompt comprising four components:
\vspace{-3pt}
\begin{itemize}[leftmargin=*]
    \item An instruction, formulated to guide the LLM in generating textual content by role-playing ~\cite{sun2023principled,white2023prompt} as a tutorial authoring expert. 
    \item Description of each textual content, to specify the requirements for the content to be generated. 
    \item Few-shot  manually-crafted demonstrations~\cite{sahoo2024systematic}, which can assist LLMs in better understanding the task and requirements described in the instruction.
    \item Task description and the filtered step sequence log, which serve as the raw information for LLMs to generate textual content.
\end{itemize}
\vspace{-3pt}

To ensure a well-structured format, we define a JSON template and specify a structured output format for LLM calls. We then align generated step titles and descriptions with the corresponding visually enhanced screenshots by step index. Finally, we complete the preparation of textual and visual materials (using file paths) by integrating predefined introductions, conclusions, initial/final state screenshots, and background images into the JSON file.

\vspace{-3pt}
\subsubsection{Tutorial Document Synthesis} 
We provide the document constructor with a JSON file containing textual content and image paths. The document constructor then generates tutorials in both HTML and Markdown formats. Following a predefined format scheme, the constructor places the title at the top center of the document as the heading. It subsequently traverses the JSON file to process the introduction, operation steps, and the output presentation module, preserving the logical order and hierarchical structure of the tutorial to ensure that users can quickly grasp the overall workflow. For each module, the corresponding data are formatted into a block containing a title, an image, and the description, thereby conveying information through both text and visuals. Finally, these blocks are sequentially integrated into the main body of the document.

 % Following a predefined format scheme, the constructor places the title at the top center and then traverses the JSON file to process the introduction, operation steps, and output presentation module while preserving the tutorial’s logical and hierarchical structure for efficient workflow understanding. Each module is formatted as a block with a title, image, and description, and these blocks are sequentially integrated into the document body.

\vspace{-3pt}
\subsubsection{Tutorial Video Synthesis}
Upon receiving a JSON file containing multi-modal resources, the video constructor generates and integrates the following segments:
(1) Introduction. The video title, initial background, and task overview are combined to form the introductory segment, enabling users to quickly understand the tutorial’s theme.
(2) Initial state presentation. The initial screenshot is displayed together with predefined explanatory text, representing the starting point of the task.
(3) Step-by-step demonstration. For each operation step, the constructor iteratively processes the content, including the step index, step title, annotated action image, and step description.
(4) Final state presentation. Upon task completion, the final state screenshots are displayed, indicating that they represent the completed outcome. 
(5) Ending segment. Following common practices in instructional videos, the tutorial concludes with a message thanking users for watching.
% \begin{enumerate}[leftmargin=*]
% \item Introduction. The video title, initial background, and task overview are combined to form the introductory segment, enabling users to quickly understand the tutorial’s theme.
% \item Initial state presentation. The initial screenshot is displayed together with predefined explanatory text, representing the starting point of the task.
% \item Step-by-step demonstration. For each operation step, the constructor iteratively processes the content, including the step index, step title, annotated action image, and step description.
% \item Final state presentation. Upon task completion, the final state screenshots are displayed, indicating that they represent the completed outcome. 
% \item Ending segment. Following common practices in instructional videos, the tutorial concludes with a message thanking users for watching.
% \end{enumerate}

In addition, all video segments are accompanied by audio narration. Subtitles and audio are precisely synchronized to ensure audiovisual consistency, thereby providing an engaging learning experience ~\cite{brame2017effective,kay2014developing}. From a visual design perspective, we adopt adaptive typography to ensure that video titles of varying lengths are evenly distributed across line widths, producing a harmonious presentation that satisfies aesthetic constraints. In addition, we carefully design the overall tutorial layout to maintain a clean and organized structure, thereby preventing learners from becoming overwhelmed by redundant or densely packed information~\cite{kay2014developing,clark2011learning}.

%% file: sections/04_experiments.tex
\vspace{-5pt}
\section{Experiments}

% highlight 每一部分用了哪个模型。
% apply到了同一个agent+Generation是同一组setting
% rq 23 合并
% 把目前的metric分散到不同的rq中
% 简单介绍一下baseline

To comprehensively evaluate the performance of our automated tutorial generation framework, we designed a series of experiments to address three key Research Questions (RQs):
\begin{itemize}[leftmargin=*]
    \item \textbf{RQ1:} How does ExeAgent perform compared to existing CUAs in terms of success rate?
    \item \textbf{RQ2:} Does our framework generate high-quality Excel tutorials? Do human evaluations and LLM-based evaluations yield consistent results? What are the quality differences among Excel tutorials generated by different LLMs? How do auto-generated tutorials compare with those authored by domain experts?
    \item \textbf{RQ3:} What are the costs (time, money, tokens, steps) of automatically generating tutorials?
    % \cz{I feel RQ23 can be merged.}
\end{itemize}

\vspace{-3pt}
\subsection{Experiment Setup} 

\subsubsection{Baseline and Backbone Models} 
To evaluate the performance of our framework, we apply it to three state-of-the-art LLMs: GPT-4.1~\cite{openai2025gpt41}, GPT-o3~\cite{openai2025o3o4} and GPT-5~\cite{openai2025gpt5}. Access to all models is provided via their respective official APIs. The specific model versions used are gpt-4.1-2025-04-14, gpt-o3-2025-04-16 and gpt-5-2025-08-07. We use the same LLM as the backbone model for both the ExeAgent and tutorial generation.

\vspace{-3pt}
\subsubsection{Parameter Setting} 
For LLM model configurations, we set the temperature to 0.01, top\_p to 0.95, and the max\_tokens parameter to 4096 for each model, while other parameters are maintained at their default settings as specified in the API documentation~\cite{openai2024api}.

\vspace{-3pt}
\subsection{RQ1: Success Rate and Performance }
\label{subsec:rq1,Success Rate and Performance Comparison}

We compare the success rate with state-of-the-art GUI agents on the dataset in \autoref{subsec:dataset contruction}. As \autoref{tab:success rates comparison} shows, ExeAgent achieves a 39.58\% success rate across the benchmark, surpassing the best-performing baseline (UFO) by 8.47\%. Also, compared to other CUAs, ExeAgent accomplishes tasks with fewer steps, thereby demonstrating higher efficiency. When using GPT-4.1 as the base model, ExeAgent completes tasks in an average of 4.57 steps, which is substantially lower than UFO (5.79 steps) and Operator (8.38 steps). 
% Furthermore, compared to other CUAs, ExeAgent completes tasks with fewer steps, showcasing its efficiency, whereas UFO and Operator tend to generate longer step sequences that are less effective. This difference arises because, when an operation fails, these frameworks often repeat the same action until it succeeds or alter the execution plan.
% In contrast, ExeAgent leverages APIs to perform operations with higher precision, avoiding ineffective steps and completing tasks with minimal actions.

These results underscore the effectiveness of ExeAgent in completing complex Excel-specific tasks through API-assisted execution. This approach allows accurate execution of Excel operations, reduces vulnerability caused by GUI variations, and significantly enhances the reliability of multi-step workflows. In comparison, the performance of baselines is primarily attributable to inconsistencies between planned and executed actions, such as selecting the wrong control or performing unintended operations. Execution errors often stem from inaccurate visual reasoning, incorrect associations between GUI elements and actions, or erroneous inference by the LLM. Unlike tasks dominated by simple click-based operations on web pages or other desktop applications, Excel tasks involve precise selections, drag-and-drop actions, and other high-precision operations. The intricate and complex nature of Excel operations increases the possibilities of such errors. By integrating API calls in execution, ExeAgent can accurately execute specific actions via parameter passing, preventing errors during task execution.

Interestingly, although GPT-5 exhibits strong general capabilities, its success rate is not outstanding. We found that it is because GPT-5 tends to adopt an overly ``comprehensive'' approach when reasoning about a task, resulting in unnecessary operations, redundant execution steps, and an increased probability of errors and evaluation difficulty. For example, when selecting a table range, GPT-5 enforces explicit confirmation of the active spreadsheet, thereby introducing an additional step for sheet selection.

\begin{table}[h!]
\vspace{-7pt}
\centering
\caption{Success rates (\%) and step number across agents.}
\vspace{-10pt}
{\setlength{\tabcolsep}{2pt} 
\begin{tabular}{llccccc}
\toprule
Agent & Model & Search & Websites & In-App & All & \# Step\\
\midrule

UFO & GPT-4.1 & 30.00& 32.73& 28.26 & 31.11 &5.79\\
Operator & computer-use & 28.71& 24.83& 25.36& 26.62 & 8.38\\
% UFO\textsuperscript{2} & GPT-4.1 & 35.71 & 26.46& 32.61 & 31.17\\
ExeAgent & GPT-5 & 26.43 & 30.79 & 26.09 & 28.42 & 5.51\\
ExeAgent & GPT-o3 & 28.43 & 39.67 & 33.33 & 34.06 &5.72\\
ExeAgent & GPT-4.1 & \textbf{39.29} & \textbf{40.78} & \textbf{34.78} & \textbf{39.58} & \textbf{4.57}\\
\bottomrule
\end{tabular}
}
\label{tab:success rates comparison}
\vspace{-11pt}
\end{table}

\vspace{-3pt}
\subsection{RQ2: Quality of Generated Tutorials}

\begin{table*}[t]
\vspace{-3pt}
\centering
\caption{Mean of human and LLM ratings across document metrics.}
\vspace{-11pt}
\label{tab:doc_metrics_score_with_def}
\setlength{\tabcolsep}{3pt} % 稍微收紧列间距以适应更多内容
\begin{tabularx}{0.95\textwidth}{l@{\hspace{12pt}} X cccccc}
\toprule
\multicolumn{1}{c}{\multirow{2}{*}{\textbf{Metric}}} & 
\multicolumn{1}{c}{\multirow{2}{*}{\textbf{Definition}}} 
& \multicolumn{2}{c}{\textbf{GPT-4.1}} 
& \multicolumn{2}{c}{\textbf{GPT-o3}} 
& \multicolumn{2}{c}{\textbf{GPT-5}} \\
\cmidrule(lr){3-4} \cmidrule(lr){5-6} \cmidrule(lr){7-8}
& & \textbf{Human} & \textbf{LLM} & \textbf{Human} & \textbf{LLM} & \textbf{Human} & \textbf{LLM} \\
\midrule
Text--Image Mapping & Text and images are accurately aligned. & 4.75 & 4.98 & 4.73 & 4.98 & 4.63 & 4.98 \\
Clarity & Each step is described explicitly and unambiguously. & 4.69 & 4.90 & 4.66 & 4.88 & 4.56 & 4.84 \\
Completeness & All essential operations are covered. & 4.68 & 4.98 & 4.59 & 4.96 & 4.73 & 4.92 \\
Correctness & All necessary steps are performed accurately. & 4.67 & 4.96 & 4.59 & 4.96 & 4.58 & 4.94 \\
Sequential Order & Steps are presented in a logical sequence. & 4.63 & 4.92 & 4.62 & 4.92 & 4.76 & 4.90 \\
Understandability & The content is intuitive and easy to follow. & 4.63 & 4.88 & 4.60 & 4.66 & 4.61 & 4.90 \\
Conciseness & Redundant or irrelevant steps are avoided. & 4.56 & 4.72 & 4.51 & 4.70 & 4.43 & 4.56 \\
Task Completion & Users can complete the tasks smoothly. & 4.53 & 4.74 & 4.41 & 4.76 & 4.52 & 4.90 \\
Efficiency & Format enables faster task completion. & 4.38 & 4.48 & 4.24 & 4.44 & 4.31 & 4.56 \\
Satisfaction & Users are satisfied with the tutorial. & 4.29 & 4.58 & 4.12 & 4.56 & 4.32 & 4.80 \\
Preference & Users prefer this type over other formats. & 4.24 & 4.58 & 4.09 & 4.46 & 4.27 & 4.62 \\
\midrule
\textbf{Average} & Average of all document metrics. & \textbf{4.55} & \textbf{4.79} & \textbf{4.47} & \textbf{4.75} & \textbf{4.52} & \textbf{4.81} \\
\bottomrule
\end{tabularx}
\vspace{-5pt}
\end{table*}

\begin{table*}[t]
\centering
\caption{Mean of human and LLM ratings across video metrics.}
\vspace{-10pt}
\label{tab:video_metrics_score_with_def}
\setlength{\tabcolsep}{1pt} % 调整列间距
\begin{tabularx}{1\textwidth}{l@{\hspace{10pt}} X c c c c c c}
\toprule
\multicolumn{1}{c}{\multirow{2}{*}{\textbf{Metric}}} & 
\multicolumn{1}{c}{\multirow{2}{*}{\textbf{Definition}}} 
& \multicolumn{2}{c}{\textbf{GPT-4.1}} 
& \multicolumn{2}{c}{\textbf{GPT-o3}} 
& \multicolumn{2}{c}{\textbf{GPT-5}} \\
\cmidrule(lr){3-4} \cmidrule(lr){5-6} \cmidrule(lr){7-8}
& & \textbf{Human} & \textbf{LLM} & \textbf{Human} & \textbf{LLM} & \textbf{Human} & \textbf{LLM} \\
\midrule
Design Quality & The video is well-structured and user-friendly. & 4.80 & 4.40 & 4.66 & 4.28 & 4.68 & 4.44 \\
Interactivity & Key operations are highlighted, allowing users to easily notice them. & 4.80 & 4.90 & 4.62 & 4.82 & 4.64 & 4.84 \\
Transferability & The methods can be effectively applied to similar tasks. & 4.70 & 4.68 & 4.51 & 4.68 & 4.55 & 4.62 \\
Usability & The operation process is clear, enabling users to follow it easily. & 4.55 & 4.74 & 4.40 & 4.82 & 4.44 & 4.82 \\
Comp \& Sat & Users can complete the task smoothly, and their overall satisfaction is high. & 4.53 & 4.74 & 4.47 & 4.86 & 4.52 & 4.86 \\
Correctness & All necessary steps are performed completely and accurately. & 4.50 & 4.90 & 4.61 & 4.94 & 4.59 & 4.92 \\
Eff \& Pref & The tutorial improves efficiency and is the preferred format for future learning. & 4.50 & 4.70 & 4.32 & 4.82 & 4.38 & 4.78 \\
\midrule
\textbf{Average} & Average of all video metrics. & \textbf{4.62} & \textbf{4.73} & \textbf{4.51} & \textbf{4.75} & \textbf{4.54} & \textbf{4.75} \\
\bottomrule
\end{tabularx}
\vspace{-8pt}
\end{table*}

\subsubsection{Evaluation Metrics} 
% \cz{List them?}
\label{sec:eval_metrics}
% To comprehensively evaluate the quality of our generated Excel tutorials, we define evaluation metrics for documents and videos separately. 

To rigorously assess the quality of the generated Excel tutorials, we establish distinct evaluation metrics for tutorial documents and videos.
Building upon prior work ~\cite{zhong2021helpviz, liu2024having,morain2012yoututorial, swarts2012new, yousef2014drives} and tailoring it to the specific characteristics of our task, we establish 11 evaluation metrics for the documents (\autoref{tab:doc_metrics_score_with_def}) and 7 evaluation metrics for the videos (\autoref{tab:video_metrics_score_with_def}).
% We investigate existing research~\cite{zhong2021helpviz,liu2024having,morain2012yoututorial,swarts2012new,yousef2014drives} on automated document and video generation to collect metrics used by researchers. After gathering all metrics, we merge those with similar semantics into a single metric to avoid redundancy~\cite{li2024only}. 

We design the evaluation metrics from both objective and subjective perspectives. On the objective side, the focus lies on assessing whether the tutorial description is expressed with clarity and accuracy~\cite{kay2014developing,yousef2014drives}, whether the tutorial content is complete and correct~\cite{yousef2014drives}, whether the structure is logically coherent~\cite{zhong2021helpviz,liu2024having}, and whether the alignment between text and visuals is consistent~\cite{brame2017effective,van2013eight}. These criteria ensure that the material provides reliable, understandable, and executable guidance for users. 
On the subjective side, the metrics emphasize user experience~\cite{sauro2016quantifying}, including content comprehensibility, execution efficiency, task completion, as well as user satisfaction and preference~\cite{zhong2021helpviz,liu2024having}. These criteria ensure that the tutorials can meet users’ practical needs and expectations. This dual design not only guarantees the scientific rigor and objectivity of the evaluation but also reflects the practical experiences of users.

\vspace{-5pt}
\subsubsection{Human Evaluation}
\label{subsec:human_evaluation_details} 

% recruit volunteer xx， 有difference experience。 diversity比较重要。每个人打多少个样本，每个样本被几个人打分。选了哪些metric

For the human evaluation, we recruited 10 participants with diverse levels of Excel experience via online platforms, all of whom were unaffiliated with the project, to independently evaluate the automatically generated tutorials. From the full set of generated tutorials, we selected a representative sample of 50 tasks. To achieve diversity among the selected cases, we ensure that the 50 tasks cover all operation categories and target object categories through rejection sampling. In total, we evaluate the Excel tutorial documents and videos generated by the three best-performing LLMs (GPT-4.1, GPT-o3, GPT-5), covering a total of 300 tutorials (150 tutorial documents and 150 videos).

We ensure each tutorial is rated by at least 2 evaluators independently. The final score was calculated as the average of the two evaluators’ ratings. Participants were instructed to rate the assigned cases based on the criteria in ~\autoref{sec:eval_metrics}, using a 5-point Likert scale with explicit descriptions for each score. The average ratings for each case were used as the final evaluation result.

In addition, participants were asked to complete an open-ended questionnaire designed to capture  background information and subjective feedback, including: (1) basic demographic and background information, such as age, years of Excel usage, and self-reported proficiency level; (2) perceptions of the strengths and weaknesses of the automatically generated instructional documents and videos; (3) we conducted comparative evaluations of the automatically generated content against expert-authored materials, {where we provided both types of tutorials for the same 20 tasks and asked participants to compare them during actual use. The responses to this questionnaire provide complementary qualitative insights into participants’ perceptions, thereby informing future improvements to the automatic content generation system.
% The responses to this questionnaire provide complementary qualitative insights into participants’ perceptions, thereby informing future improvements to the automatic content generation system.

The evaluation results are presented in \autoref{tab:doc_metrics_score_with_def} and \autoref{tab:video_metrics_score_with_def}. 
Overall, the average ratings for all metrics across the three base model settings exceeded 4. This suggests that participants gave positive evaluations for the quality of documents and video tutorials generated by different models across all 18 assessment metrics. In addition, the low average variance (0.28 - 0.45) indicates that participants’ ratings are consistently concentrated across all tutorials.

\textbf{Tutorial Document Rating Analysis.} Six metrics (Clarity, Completeness, Correctness, Sequential Order, Understandability, and Text-Image Mapping) got high average ratings, which exceeded 4.5 across all models. 
These results indicate that the generated document tutorials are accurate, well-structured, and complete. This further demonstrates that ExeAgent achieves high execution fidelity and produces correct and comprehensive operational sequences.
In addition, the strong performance on Understandability and Text-Image Mapping suggests that the tutorials are highly interpretable, featuring clear textual descriptions and precisely annotated images that effectively highlight key instructional steps.
Two additional metrics, Conciseness and Task Completion, also achieved average ratings above 4.5 for most models. However, for GPT-5, the Conciseness rating was slightly lower at 4.31, indicating that while the generated tutorials are generally concise, some redundancy remains. This observation aligns with the findings discussed in \autoref{subsec:rq1,Success Rate and Performance Comparison}. The redundancy may stem from ExeAgent (GPT-5) occasionally performing exploratory or repetitive actions during task execution, which can reduce conciseness and require users to exert additional effort to complete the tasks.
Scores for Efficiency, Satisfaction, and Preference ranged from 4.0 to 4.5 across all models. These results indicate that users generally completed tasks efficiently, were satisfied with the tutorials, and preferred using them, though occasional redundancy or errors caused minor confusion.

\begin{figure*}[htbp]
    \vspace{-5pt}
    \centering
    \includegraphics[width=0.96\textwidth]{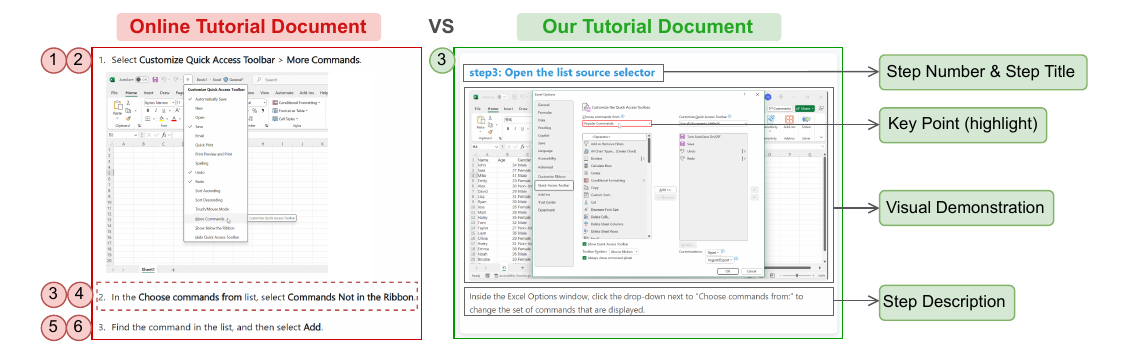}
    \vspace{-5pt}
    \caption{Comparison between our automatically generated tutorial and an expert-authored online tutorial for the task ``Add the tool to the Quick Access Toolbar in Excel''. Content highlighted in green corresponds to Step 3 of our tutorial, while content enclosed in red rectangles represents the full online tutorial. Circled numbers indicate the order of the tutorial steps.}
    \label{fig:case_study}
    \vspace{-8pt}
\end{figure*}

\textbf{Tutorial Video Rating Analysis.} 
Four metrics (Design Quality, Interactivity, Transferability, and Correctness) achieved average ratings exceeding 4.5 across all models. These results suggest that the generated video tutorials were perceived by participants as accurate, engaging, well-structured, and broadly applicable. By integrating both API and GUI based approaches, ExeAgent ensures the correctness of operational sequences while enhancing user attention through visual cues, thereby improving interactivity. Furthermore, the video generation process emphasizes aesthetic presentation and the generalization of tutorial content.
The Completion and Satisfaction (Comp \& Sat) metric averaged around 4.5, indicating that most video tutorials effectively guided users through task completion and yielded a satisfactory experience. However, in some cases, insufficient clarity in step-level details may have required users to engage in self-reflection during execution.
Scores for Efficiency and Preference (Eff \& Pref) ranged between 4.0 and 4.5 across models, suggesting that the tutorials generally improved task efficiency and were favorably received by users. Nonetheless, compared to expert-produced tutorials, the generated videos may lack vividness and operational fluency due to occasional redundancy, which could negatively impact the overall user experience.

\vspace{-5pt}
\subsubsection{LLM Evaluation} 
For the automated evaluation, we employed GPT-4.1 to rate the generated tutorials across all metrics using a 5-point Likert scale. The rating criteria for the LLM evaluation were identical to those used in the human evaluation. To facilitate comprehension by the model, each metric was accompanied by detailed explanations, and two concrete examples were provided to illustrate the expected ratings. Also, the model was instructed to output both the assigned score and a justification for that score.

The evaluation results are presented in \autoref{tab:doc_metrics_score_with_def} and \autoref{tab:video_metrics_score_with_def}. LLM ratings are predominantly above 4.5 for both documents and videos, indicating that LLMs express a high level of approval for the generated tutorials. 
Moreover, except for Design Quality, LLM ratings exceed the corresponding human ratings.
The relatively lower ratings in Design Quality may be attributed to the weakness of current LLMs to process audio information, which limits their capacity to evaluate the alignment between audio, textual, and visual elements.
Consequently, LLMs tend to assign a more conservative rating of 4 rather than 5 for this metric. 
Other metrics with relatively lower LLM ratings are concentrated in more subjective aspects, such as Task Efficiency, Satisfaction, and Preference. One possible reason is that, compared with objective metrics, LLMs are less confident in evaluating subjective dimensions, resulting in a tendency to assign non-maximal ratings.

\vspace{-5pt}
\subsubsection{Evaluation Summary}
Overall, the tutorials generated in this study received high ratings from both human assessors and LLM. The average rating from humans was approximately 4.5, while the mean LLM rating reached 4.7. This result indicates that the generated tutorials achieved a high standard in both objective accuracy and subjective user experience. Notably, both the average rating and the proportion of high ratings
% (see \autoref{fig:score_distribution})
from the LLM were higher than those from the human assessors. 
We hypothesize that this discrepancy may arise from the LLM's limitations in identifying subtle flaws within the content and in perceiving nuanced human subjective feedback, leading to a tendency to award higher ratings.

Furthermore, both document and video tutorials exhibit positive correlations and a high level of agreement between LLM-based and human evaluations. Across the 18 evaluation metrics, the Pearson correlation coefficients range from 0.13 to 0.44, and the mean absolute differences between LLM and human ratings range from 0.27 to 0.68. In addition, from a pool of 300 tutorials scored by the LLM, we randomly sampled 50 video tutorials and 50 document tutorials (resulting in 900 scores across 18 metrics) and invited three Excel experts to independently assess the LLM-assigned scores and their accompanying justifications. The results show that in 92.9\% of the cases, the LLM scores and justifications were judged to be correct. These findings indicate consistency between LLM and human evaluations, supporting the high reliability of LLM-based scoring.

%原来的
% Furthermore, we investigate the correlations between LLM-based evaluations and human assessments using both Pearson and Kendall metrics
% % , as shown in \autoref{tab:overall_correlation}
% . Overall, both document and video tutorials exhibit positive correlations, with average correlation coefficients ranging between 0.21 and 0.44. Notably, the correlations for documents are consistently higher than those for videos. This discrepancy may be attributed to the inherent limitations of current LLMs in comprehensively understanding multi-modal content—particularly tutorials that integrate textual descriptions, visual elements, and audio narration. As LLMs lack the capability to process audio and fully interpret visual context, their evaluations may not accurately reflect the holistic quality of video tutorials.

\vspace{-3pt}
\subsubsection{User Study on Expert-Authored Tutorials}
As described in \autoref{subsec:human_evaluation_details}, ten participants were asked to respond to four open-ended questions regarding the advantages and limitations of our tutorials compared to expert-authored ones.

\textbf{Document Tutorials.}
Participants generally found our document tutorials to be more detailed and intuitive. They highlighted the step-by-step guidance supported by screenshots and annotated highlights as particularly helpful. For instance, Participant 1 (P1) noted that many expert-authored tutorials lacked visual illustrations, making them difficult to follow, whereas our tutorials consistently included images for each step. P7 and P8 emphasized that our tutorials explained every step thoroughly, while expert-authored documents often assumed prior knowledge and used overly concise descriptions, which hindered sequential understanding. Additionally, P3 remarked that our tutorials provided explanatory content that not only supported task completion but also conveyed underlying principles of Excel, enhancing knowledge transferability.

% \begin{table*}[t]
% \centering
% \caption{Performance across different software}
% \vspace{-8pt}
% \label{tab:software_results}
% \renewcommand{\arraystretch}{0.9}
% \begin{tabular}{lcccccccc}
% \toprule
% \multirow{2}{*}{Software} 
% & \multirow{2}{*}{Acc (\%)} 
% & \multirow{2}{*}{\# Step} 
% & \multirow{2}{*}{Money (\$)} 
% & \multirow{2}{*}{Time (s)} 
% & \multicolumn{2}{c}{Document score} 
% & \multicolumn{2}{c}{Video score} \\
% \cmidrule(lr){6-7} \cmidrule(lr){8-9}
% & & & & & Human & LLM & Human & LLM \\
% \midrule
% Word & 61.78 & 4.16 & 0.26 & 253.28 & 4.63 & 4.68 & 4.61 & 4.55 \\
% PowerPoint  & 63.13 & 4.12 & 0.25 & 249.93 & 4.71 & 4.73 & 4.69 & 4.62 \\
% \bottomrule
% \end{tabular}
% \vspace{-5pt}
% \end{table*}

\textbf{Video Tutorials.}
Participants appreciated the clearer segmentation of steps, moderate pacing, and the inclusion of subtitles in our video tutorials. P2 and P3 observed that many expert-authored videos were simple screen recordings without explicit step boundaries, whereas our videos followed a structured, step-by-step format. P9 noted that expert videos were often fast-paced and lacked subtitles, requiring frequent pauses for comprehension. In contrast, our videos maintained an appropriate pace and included subtitles, which facilitated both understanding and task replication.

\textbf{Suggestions for Improvement.}
Participants also identified areas for enhancement. In addition to using highlights, they suggested incorporating magnification to better emphasize operational regions. Furthermore, they recommended enriching the tutorials with contextual background, such as application scenarios of the features, to improve content depth and relevance, drawing inspiration from expert-authored materials.

\vspace{-3pt}
\subsubsection{Case Study}
We conduct a comparative analysis of online expert-authored tutorials and our automatically generated tutorials on the task ``add the tool to the Quick Access Toolbar in Excel''.

Our tutorials decompose the task into seven explicit atomic steps: 
(1) Open the quick access toolbar customization menu.
(2) Access more commands for customization.
(3) Open the quick access toolbar options.
(4) Switch to Commands Not in the Ribbon.
(5) Select the tool.
(6) Add the tool to the quick access toolbar.
(7) Confirm and apply the customization.
Each step is clearly delineated and accompanied by a step number, a title, a detailed description, a screenshot, and a highlighted marker indicating the exact operation location (see \autoref{fig:case_study}). In addition, both the initial and completed states of the file are provided as visual references. The tutorial video further integrates narration with introductory and concluding slides to enhance accessibility and comprehension. 

In contrast, expert-authored tutorials have several limitations (see \autoref{fig:case_study}). Owing to the authors’ high familiarity with Excel, certain steps in existing help documents are either omitted or merged. Furthermore, many documents lack visual aids and explicit highlighting of operation areas, creating comprehension difficulties for users. For example, Steps 1--2, 3--4, and 5--6 are often collapsed into a single step in expert-authored tutorials, while Step~7 is omitted entirely. In terms of illustrations, only a few steps are occasionally supported by a screenshot, while the remaining steps are left without visual guidance. Additionally, the highlighted markers for key operations are often omitted. As a result, novice users without prior background knowledge may struggle to follow the procedure and complete tasks efficiently. Similarly, most expert-authored help videos consist of fast-paced screen recordings that neither explicitly separate steps nor provide subtitles or textual annotations, relying solely on narration. As a result, the videos are difficult for viewers to follow and offer reduced clarity and accessibility.

% Similarly, most expert-authored help videos are simple screen recordings that do not explicitly name or separate individual steps. The fast-paced presentation makes videos difficult for viewers to follow along effectively. Moreover, existing videos typically lack subtitles or textual annotations, relying solely on narration, which reduces intuitiveness and limits accessibility.

% For manually authored video tutorials, most existing ones are presented as continuous screen recordings, lacking step-by-step segmentation and clear structural organization. 
% Moreover, such tutorials exhibit common deficiencies, including: (1) an absence of subtitles; (2) a lack of voice-over narration; and (3) a failure to highlight key operational areas on the screen.

\vspace{-5pt}
\subsection{RQ3: Cost Analysis of Tutorial Generation}

\begin{table}[!t]
\vspace{-1pt}
\centering
\caption{Tutorial Generation Cost.}
\vspace{-8pt}

\label{tab:cost}
\begin{tabular}{l c c c c}
\toprule
Model & \# Step & Token (k) & Money (\$) & Time (s) \\
\midrule
GPT-4.1 & \textbf{4.57} & \textbf{132.15} & \textbf{0.28} & \textbf{270.58} \\
GPT-o3  & 5.72 & 183.24 & 2.00 & 431.15 \\
GPT-5   & 5.51 & 142.86 & 0.48 & 408.93 \\
\bottomrule
\end{tabular}
\vspace{-10pt}
\end{table}

As shown in \hyperref[tab:cost]{Table~\ref*{tab:cost}}, the three models exhibit differences in cost-efficiency when generating tutorials, primarily due to their underlying architectures and training paradigms. GPT-4.1, as a generative model, achieves the highest efficiency across all evaluated dimensions (step count, token usage, execution time, and monetary cost). This advantage stems from its optimization for direct instruction execution. For tasks such as Excel tutorial generation, which rely heavily on existing knowledge, GPT-4.1 can produce high-quality action sequences without redundant intermediate reasoning or reflective steps, thereby yielding the lowest overall cost. In contrast, GPT-o3, designed as a reasoning model, incurs the highest resource consumption. Its training paradigm requires the model to ``think'' explicitly before producing a final answer, generating multiple intermediate reasoning traces and performing reflective checks, which substantially increases cost. GPT-5 integrates both generative and reasoning paradigms, with cost metrics falling between those of GPT-4.1 and GPT-o3, reflecting a trade-off between generation efficiency and reasoning depth. Overall, for the specific application of Excel tutorial generation, GPT-4.1’s direct generation strategy offers clear advantages, making it the most cost-effective model.

Additionally, prior studies have emphasized the substantial cost of manually producing tutorials~\cite{kim2014crowdsourcing,hu2023smartrecorder}. To further quantify this cost, we asked three Excel experts to manually generate documentation and video tutorials while recording the time required. On average, manual tutorial creation required approximately 2.5 hours per task. In contrast, our automated approach completes the same process in only \textbf{1/20} of the time, greatly reducing human effort.

%% file: sections/06_discussion.tex
\vspace{-5pt}

\section{Deployment \& Discussion}

\subsection{Production Impact}

The proposed framework has been deployed as a core production component within Microsoft M365 to automatically generate in-product help and tutorial content at scale. Using our framework, we generated approximately 1,400 Excel help articles within a three-week period. In contrast, producing the same volume of content through manual authoring would have required over 300 person-weeks of effort. This corresponds to a time reduction of more than \textbf{99\%}, demonstrating a substantial improvement in production efficiency and content throughput.

Beyond efficiency gains, the generated tutorials were released to real users in production. Online evaluation shows that the new content achieved a \textbf{9\%} increase in user engagement compared to previously deployed, human-authored tutorials. This improvement indicates that the automatically generated content not only meets production-quality standards, but also delivers superior usefulness and clarity from an end-user perspective.

Overall, these results demonstrate that our framework is not merely a research prototype, but a robust and scalable framework that delivers tangible impact in a real-world production environment. It significantly reduces authoring cost and latency while improving user engagement, making it a practical and effective solution for large-scale tutorial and help content generation in modern software systems.

\subsection{Framework Generalizability}

\begin{table}[t]
\vspace{-1pt}
\centering
\caption{Performance across different software applications.}
\vspace{-8pt}
\label{tab:software_results}
\renewcommand{\arraystretch}{0.9}
\setlength{\tabcolsep}{2pt}
\begin{tabular}{lcccccccc}
\toprule
\multirow{2}{*}{\shortstack{Soft-\\ware}}
& \multirow{2}{*}{\shortstack{Acc\\(\%)}} 
& \multirow{2}{*}{\shortstack{\# Step}} 
& \multirow{2}{*}{\shortstack{Money\\(\$)}} 
& \multirow{2}{*}{\shortstack{Time\\(s)}} 
& \multicolumn{2}{c}{Document} 
& \multicolumn{2}{c}{Video} \\
\cmidrule(lr){6-7} \cmidrule(lr){8-9}
& & & & & Human & LLM & Human & LLM \\
\midrule
Word & 61.78 & 4.16 & 0.26 & 253.28 & 4.63 & 4.68 & 4.61 & 4.55 \\
PPT  & 63.13 & 4.12 & 0.25 & 249.93 & 4.71 & 4.73 & 4.69 & 4.62 \\
\bottomrule
\end{tabular}
\vspace{-12pt}
\end{table}

Our framework is readily generalizable to other software applications since its core modules are software-agnostic and can be directly reused. Adapting to a new application requires only collecting the corresponding data templates (tasks and files) and integrating the software-specific APIs into the automatic trace collection module (Execution Agent). To evaluate this generalizability, we collected user queries for Word and PowerPoint from real-world sources (search engines, community forums, in-application help content) and randomly sampled 500 queries per application for testing. We conducted tutorial generation testing using GPT-4.1, which achieved the highest success rate in our Excel experiments. As shown in \hyperref[tab:software_results]{Table~\ref*{tab:software_results}}, our framework generalizes well and can be easily extended to other software applications.

%% file: sections/07_conclusion.tex
\vspace{-5pt}
\section{Conclusion}

In this paper, we formalize the problem of task-to-tutorial generation, develop an evaluation protocol covering eighteen key aspects, and present a dataset comprising 1,559 diverse Excel tasks. To address this problem, we propose the first framework capable of automatically generating Excel tutorials solely from task descriptions. Our framework adopts a three-stage strategy consisting of (1) task instantiation, (2) automatic trajectory collection, and (3) tutorial generation. Experimental results demonstrate that our ExeAgent significantly outperforms baseline methods in terms of the success rate of Excel task completion, achieving an improvement of 8.5\%. Both human evaluation and LLM evaluation confirm the high quality of the generated tutorials. Furthermore, a case study indicates that our tutorials match or occasionally exceed the quality of expert-authored materials. Finally, cost analysis and generalizability evaluations reveal that our approach substantially reduces both time and monetary expenses compared with manual tutorial creation, while also being readily extensible to different softwares.

% \section*{Data Availability}
% The data that support the findings of this study are available on request from the corresponding author, upon reasonable request.